\documentclass[12pt]{elsarticle}
\usepackage{graphicx}
\usepackage{color}
\usepackage{latexsym}
\usepackage{amsmath}
\usepackage{amssymb}

\usepackage{bbm}                                                  

\def\nc{\newcommand}
\nc{\beq}{\begin{equation}}  \nc{\eeq}{\end{equation}}
\nc{\bea}{\begin{eqnarray}}  \nc{\eea}{\end{eqnarray}}
\nc{\baa}{\begin{array}}     \nc{\eaa}{\end{array}}
\nc{\bit}{\begin{itemize}}   \nc{\eit}{\end{itemize}}
\nc{\ben}{\begin{enumerate}} \nc{\een}{\end{enumerate}}
\nc{\bce}{\begin{center}}    \nc{\ece}{\end{center}}
\nc{\bpm}{\begin{pmatrix}}   \nc{\epm}{\end{pmatrix}}
\nc{\bvt}{\begin{verbatim}}  \nc{\evt}{\end{verbatim}}
\def\half{\frac12}	

\def\to{\rightarrow}

\def\gesim{\gtrsim}
\def\lesim{\lesssim}
\def\gesim{\,{\raise-3pt\hbox{$\sim$}}\!\!\!\!\!{\raise2pt\hbox{$>$}}\,}
\def\lesim{\,{\raise-3pt\hbox{$\sim$}}\!\!\!\!\!{\raise2pt\hbox{$<$}}\,}
\def\boldoverdot{\,{\raise6pt\hbox{\bf.}\!\!\!\!\>}}

\def\ie{{\it i.e.}}
\def\eg{{$e.\,g.$}}

\def\etal{{\it et al.}}
\def\then{{\quad\Rightarrow\quad}}
\def\acal{{\cal A}}

\def\dcal{{\cal D}}
\def\ecal{{\cal E}}

\def\kcal{{\cal K}}
\def\lcal{{\cal L}}
\def\mcal{{\cal M}}

\def\ocal{{\cal O}}

\def\scal{{\cal S}}

\def\zcal{{\cal Z}}

\def\mati{{\mathbbm1}}

\usepackage{bm}
\def\rhs{right hand side\ }

\def\diag{\hbox{\diag}}

\def\gev{\hbox{GeV}}

\def\mn{{\mu\nu}}

\def\doubleundertext#1{
{\undertext{\vphantom{y}#1}}\par\nobreak\vskip-\the\baselineskip\vskip4pt%
\undertext{\hbox to 2in{}}}
\def\inbox#1{\vbox{\hrule\hbox{\vrule\kern5pt
     \vbox{\kern5pt#1\kern5pt}\kern5pt\vrule}\hrule}}
\def\sqr#1#2{{\vcenter{\hrule height.#2pt
      \hbox{\vrule width.#2pt height#1pt \kern#1pt
         \vrule width.#2pt}
      \hrule height.#2pt} } }
\def\square{\mathchoice\sqr56\sqr56\sqr{2.1}3\sqr{1.5}3}
\def\today{\ifcase\month\or
  January\or February\or March\or April\or May\or June\or
  July\or August\or September\or October\or November\or December\fi
  \space\number\day, \number\year}
\def\pmb#1{\setbox0=\hbox{#1}%
  \kern-.025em\copy0\kern-\wd0
  \kern.05em\copy0\kern-\wd0
  \kern-.025em\raise.0433em\box0 }

\def\pmbb#1{\setbox0=\hbox{#1}%
  \kern-.02em\copy0\kern-\wd0
  \kern.04em\copy0\kern-\wd0
  \kern-.02em\raise.03464em\box0 }
\def\su#1{{SU(#1)}}
\def\ui{U(1)}
\def\sumprime_#1{\setbox0=\hbox{$\scriptstyle{#1}$}
  \setbox2=\hbox{$\displaystyle{\sum}$}
  \setbox4=\hbox{${}'\mathsurround=0pt$}
  \dimen0=.5\wd0 \advance\dimen0 by-.5\wd2
  \ifdim\dimen0>0pt
  \ifdim\dimen0>\wd4 \kern\wd4 \else\kern\dimen0\fi\fi
\mathop{{\sum}'}_{\kern-\wd4 #1}}
\nc{\eps}{\varepsilon}
\nc{\vp}{\varphi}
\nc{\tvp}{\widetilde{\varphi}}
\nc{\D}{\mbox{$\not\!\!D$}}
\nc{\Db}{\mbox{${\raisebox{2mm}{\boldmath ${}^\leftarrow$}\hspace{-4mm} D}$}}
\nc{\Dbs}{\mbox{${\raisebox{2mm}{\boldmath ${}^\leftarrow$}\hspace{-4mm} \D}$}}
\nc{\Dfb}{\mbox{$\raisebox{2mm}{\boldmath ${}^\leftrightarrow$}\hspace{-4mm} D$}}
\nc{\Dfbs}{\mbox{$\raisebox{2mm}{\boldmath ${}^\leftrightarrow$}\hspace{-4mm} \D$}}
\nc{\vpj }{\mbox{${\vp^\dag i\,\raisebox{2mm}{\boldmath ${}^\leftrightarrow$}\hspace{-4mm} D_\mu\,\vp}$}}
\nc{\vpjt}{\mbox{${\vp^\dag i\,\raisebox{2mm}{\boldmath ${}^\leftrightarrow$}\hspace{-4mm} D_\mu^{\,I}\,\vp}$}}
\nc{\psid}{\mbox{$\overline{\psi} i\,\raisebox{2mm}{\boldmath ${}^\leftrightarrow$}\hspace{-4mm} \D\,\psi$}}
\nc{\f}{\frac}

\def\nnb{\nonumber}
\def\p{\partial}
\def\wt{\widetilde}
\def\ni{\noindent}
\def\gv{\gamma_5}
\def\mn{{\mu\nu}}

\def\ocal{Q}

\newcommand{\eqn}[1]{eq.~(\ref{#1})}

\journal{Nuclear Physics B}

\begin{document}

\begin{frontmatter}



\title{The Bases of Effective Field Theories}


\author[label1,label2]{Martin B Einhorn}
\author[label3]{Jos\'e Wudka}
\address[label1]{Kavli Institute for Theoretical Physics\footnote{Current address.}, 
University of California, Santa Barbara, CA 93106-4030}
\address[label2]{Michigan Center for Theoretical Physics, 
University of Michigan, Ann Arbor, MI 48109}
\address[label3]{Department of Physics and Astronomy, 
University of California, Riverside, CA 92521-0413}

\begin{abstract}
With reference to the equivalence theorem, we discuss the selection of
basis operators for effective field theories in general.   The
equivalence relation can be used to partition operators into equivalence
classes, from which inequivalent basis operators are selected. 
These classes can also be identified as containing
Potential-Tree-Generated (PTG) operators, Loop-Generated (LG) operators,
or both, independently of the specific dynamics of the underlying
extended models, so long as it is perturbatively decoupling.  For an
equivalence class containing both, we argue that the basis operator
should be chosen from among the PTG operators, because they may have the
largest coefficients.  We apply this classification scheme to
dimension-six operators in an illustrative Yukawa model as well in the
Standard Model (SM). We show that the basis chosen by Grzadkowski~{\it
et.~al.}~\cite{Grzadkowski:2010es} for the SM satisfies this criterion. 
In this light, we also revisit and verify our earlier
result~\cite{Arzt:1994gp} that the dimension-six corrections to the
triple-gauge-boson couplings only arise from LG operators, so the
magnitude of the coefficients should only  be a few parts per thousand
of the SM gauge coupling if BSM dynamics respects decoupling.  
The same is true of the quartic-gauge-boson couplings.
\end{abstract}


\end{frontmatter}


\pagebreak


\section{Introduction} \label{sec:intro}

Effective quantum field theories have a wide variety of applications in
condensed matter~\cite{cm} and elementary particle
physics~\cite{he1,he2}, both as methods for facilitating calculations
and as ways of exploring or constraining new physics. In this last
application, there has been a revival of interest among high energy
physicists since the discovery of a Higgs boson at the CERN LHC, 
apparently resolving the long-standing uncertainty about the theory of
elementary particles known as the Standard Model (SM).

Without knowing the precise form of new degrees of freedom or new
particles, there are inherent ambiguities in the form of additional
operators to be added to a theory because of the equivalence theorem. 
As reviewed in more detail below, this states that observable transition
amplitudes (S-matrix elements) are unchanged by replacing some operators
with others if their difference vanishes ``on-shell," \ie, if the
difference vanishes when the classical equations of motion (EoM) are
satisfied.  Although this allows one to reduce the number of new
operators and coupling constants that must be introduced
\cite{Buchmuller:1985jz, Grzadkowski:2010es}, in the face of ignorance
of the underlying dynamics, it seems the choices are both arbitrary and
irrelevant.

In any given model extending the SM to higher energy scales, some
operators $\ocal_i$ may arise from tree diagrams in an underlying
theory, while others may only emerge from loop corrections.  Generally,
the coefficients of loop diagrams, as quantum corrections to the
classical theory, are perturbatively smaller than those associated with
tree diagrams, being associated with higher powers of dimensionless
coupling constants times factors of $(16\pi^2)^{-n},$ where $n$ is the
number of loops.   Even in nonperturbative applications~\cite{he2}, such
distinctions between trees and loops can be important although less so
because of the strong interactions that are involved.  It has been
observed~\cite{Arzt:1994gp}, however, that symmetries associated with
the known dynamics, when preserved by the underlying new degrees of
freedom, may be used to classify operators arising from the underlying
dynamics, irrespective of the particular model.

Given that there is some arbitrariness in the choice of operators, there
has been a good deal of recent discussion concerning the ``best" choice
to make to perform fits to experimental data~\cite{basis, basis2}.  A number of
people have cited the difficulties deciding among  equivalent operators,
because some arise in tree-approximation while others may arise only
from loop-diagrams~\cite{Wudka:1994ny, Grzadkowski:2010es}.

In the past, it has been argued (\eg, in ref.~\cite{Wudka:1994ny}) that,
because the equivalence theorem relates some operators arising from
loops to operators arising from trees, there is no way to decide {\it a
priori} which operators to choose. In this paper, we shall discuss the
best way to choose among such higher-order operators.  The inherent
ambiguities discussed in ref.~\cite{Wudka:1994ny} can be unraveled in a
general way, independent of any particular application.  We shall limit
our discussion to perturbative applications\footnote{We will assume that
we are dealing with relativistic quantum fields in four dimensions.},
where such distinctions are most important, although perhaps this could
be extended to other applications. Elsewhere~\cite{ew}, we shall apply
this the potential influence of physics beyond the Standard Model (BSM) 
to the determination of the properties of the observed Higgs
boson.  This is what inspired the present investigation which, in the
end, led to conclusions quite independent of that motivation.

An outline of the paper is as follows:  In the next section, we review
some features of effective Lagrangians, with particular attention the
equivalence theorem.  In Appendix~A, we discuss some technical
complications associated with masses and superrenormalizable couplings.
In Section~3, we explain how the equivalence theorem can be used as an
equivalence relation to partition the set of operators. In Appendix~B,
we illustrate these concepts in a simple Yukawa model.  In Section~4, we
explain that operators may be classified as Potential-Tree-Generated
(PTG) operators or Loop-Generated (LG) operators, irrespective of the
underlying model or theory, and advocate choosing as basis vectors PTG
operators to the extent possible. In Section~5 and Appendix~C, we apply
this to the SM, and, in Section~6, we revisit our earlier
result~\cite{Arzt:1994gp} for triple-gauge-boson couplings (TGB) in this
light.  Finally, Section~7 summarizes our results.

\section{Some Features of Effective Lagrangians}\label{sec:efflag}

We begin by reviewing some of the generic properties of effective field
theories. Suppose one has a theory with a Lagrangian of the form
\beq\label{eq:model}
\lcal = \lcal'(\phi_\ell) + \Delta\lcal'(\phi_\ell,\phi_h).
\eeq
where $\lcal'$ involves ``light particles"  associated with fields
$\phi_\ell$ together with additional terms $\Delta\lcal'$ describing the
dynamics of some fields $\phi_h$ describing interactions among heavy
particles and  their couplings to light particles. These fields may
include both fermions and bosons; no distinction is necessary for
present purposes. Correspondingly, one may consider the generating
functional (or partition functional)
\beq\label{eq:action}
\zcal[j_\ell, j_h]\!=\!\!\int\! \dcal\phi_\ell \dcal\phi_h \exp\left[ i\!\int \! dx \big\{{\lcal'(\phi_\ell) + \Delta\lcal'(\phi_\ell,\phi_h)}-j_\ell\phi_\ell-j_h\phi_h\big\}\right]\!.
\eeq
At energy scales below the threshold for heavy particle production, the
collision of light particles can only produce light particles, so one
may describe their behavior  in terms of light fields alone.  The
effective action for the theory of light fields is obtained, in path
integral language, by ``integrating out" the heavy fields, or, in the
language of perturbation theory, from Feynman diagrams involving only
heavy internal propagators\footnote{There are a number of niceties suppressed 
in this formal summary.  One really wants to integrate out the heavy {\em particles}
and, to avoid having light fields create heavy particles, it is extremely convenient if 
$\Delta\lcal'$ does not contain quadratic mixing of $\phi_\ell$ with $\phi_h,$ at least
not in the kinetic terms if not in the mass terms.  This can always 
be arranged by redefining the fields and by working consistently with
the renormalized $\lcal'(\phi_\ell),$ so that $\Delta\lcal'$ contains 
counterterms that enforce this to arbitrarily high order.  Anomalous 
dimensions and $\beta$-functions may be modified accordingly.}:
\bea
\zcal[j_\ell,0]\! \! \! &=&\! \! \! \int\! \dcal\phi_\ell \exp\left[i\int dx
\left\{\lcal_{\rm eff}(\phi_\ell) - j_\ell\phi_\ell\right\}\right],\\
\exp\left[ i\!\int \! dx {\lcal_{\rm eff}(\phi_\ell)}\right]\!&\! \! \! \equiv&\! \! \! 
\exp\Big[ i\scal_{\rm eff}[\phi_\ell] \Big]\label{eq:seff}\\
&\! \! \! \equiv&\! \! \!   \exp\left[ i\!\int \! dx 
{\lcal'(\phi_\ell)}\right]\int \dcal\phi_h \exp\left[ i\!\int \! dx 
\big\{{\Delta\lcal'(\phi_\ell,\phi_h)}\big\}\right]\,\label{eq:leff}
\eea
For the integration over heavy fields in \eqn{eq:leff}), the light
fields $\phi_\ell(x)$ play the role of external sources. Perturbatively,
the integration represents the sum over all Feynman diagrams having only
virtual heavy particles, with all light particles external. In general,
the Green's functions thus obtained are nonlocal, but one may expand in
inverse powers of the heavy mass scale, which we will call $\Lambda,$ to
obtain a {\it local}\/ effective Lagrangian of the form
\beq\label{eq:efflag}
\lcal_{\rm eff}=\lcal_0(\phi_\ell) + \sum_i c_i \ocal_i(\phi_\ell),
\eeq
where $\ocal_i$ are local, Lorentz-invariant and gauge-invariant
operators of dimension greater than four with coefficients $c_i$ that
vanish as inverse powers of the heavy masses and  include
various positive powers of the coupling constants in $\Delta\lcal'$.
For simplicity, we will speak as if there is only a
single heavy scale\footnote{There is no loss of generality here, since
one could choose $\Delta \lcal'$ in eq.~(\ref{eq:model}) to be itself
an effective field theory.  One may simply ``integrate out" heavy
particles until one arrives at an effective action containing the
lightest of the heavy particles.} $\Lambda.$ The expansion also may
involve terms that grow with $\Lambda$, such as $\Lambda^2\phi_\ell^2$
or $\log(\Lambda)(\partial\phi_\ell)^2.$ Since they necessarily
involve operators of dimension four or less,  such terms can be
absorbed into the coefficients of operators in $\lcal'$ and the 
wave-function renormalization of the light
fields needed to bring the kinetic energy terms in $\lcal_0$
into canonical form\footnote{Thus, the light fields on which
$\lcal_{\rm eff}$ depend in eq.~(\ref{eq:efflag}) are {\it not} the
same as the light fields appearing in \eqn{eq:leff}.  These
redefinitions have no observable effects and are often implicit in the
literature.}. Generically, $\lcal_0$ is identical in {\it form} as
$\lcal',$ but with its fields, masses and couplings redefined.

If one knows the underlying theory, \eqn{eq:model}, then one may
calculate the coefficients $c_i$ at least to low orders in the loop
expansion.  At momentum scales below $\Lambda,$ the effective Lagrangian
\eqn{eq:efflag} may even be a more efficient method of calculation
than the original model, \eqn{eq:model}.  One familiar application
is to low-energy consequences of the SM, in which one expands in inverse
powers of the electroweak scale $v\approx 250\ \gev$ to determine weak
interaction effects on quantities such as $g_\mu-2,$ the anomalous
magnetic moment of the muon~\cite{gminus2}, weak decay amplitudes such
as beta-decay of hadrons and nuclei~\cite{Gonzalez-Alonso:2013uqa}, or
the electric dipole moment of the electron, neutron, and other particles
and atoms or  other aspects of aspects of CP-violation~\cite{cpv}.

Our current interest however concerns applications of effective field
theory to situations in which the correct underlying theory is
unknown, as when considering physics beyond the Standard Model (BSM). 
We may then characterize the entire class of decoupling models by
considering operators of successively higher-dimension, starting with
dimension-five operators, forming all possible (gauge-invariant)
operators $\ocal_i$ of a given dimension composed from the light
fields  and treating their coefficients $c_i$ as additional coupling
constants beyond those in $\lcal_0$ to be determined from or
constrained by experiment.  Since there can be many such operators, it
sounds like a daunting task to determine all these new couplings
$c_i.$  Fortunately, there number can be substantially reduced in
various ways, but in particular, by use of the equivalence
theorem which we will now review.

Operators in quantum field theory satisfy the requirements for complex
vector spaces~\cite{alg}.  
A set of operators $\{\ocal_i\}$ will be said to be {\it linearly dependent} if a linear
combination vanishes.  Stated more precisely, suppose a set
of constants (possibly depending on coupling constants) 
$\kappa_i \ne 0$ can be found such that 
\beq\label{eq:ind1}
\sum_i \kappa_i \ocal_i = 0. 
\eeq
If no such relation exists, then the set of operators $\{\ocal_i\}$
will be said to be {\it linearly independent}.  Equality here refers
only to perturbation theory, so operators that differ by a total
derivative will be considered equal, since their contributions to the
action differ only by surface terms\footnote{Thus, we ignore any
potential, topologically nontrivial terms that may arise.}.   Thus,
integration-by-parts (IBP) is allowed, \eg,
$(\partial\vp)^2+\vp\Box\vp\stackrel{IBP}{=}0.$ 
So these two operators are linearly dependent. Although not necessary,
it is often convenient, especially for renormalization, to choose the
operators $\{\ocal_i\}$ to be {\it irreducible,} in the sense that
they are contractions of monomials composed of products of the fields
and their covariant derivatives\footnote{Although $\lcal_{\rm eff}$
must be Hermitian, it is not always most expedient to make each term
in the sum Hermitian; \eg, in the SM, the Yukawa couplings to the
Higgs doublet are an example. In such cases, each term is implicitly
accompanied by its Hermitian conjugate.}.

The set of all (Lorentz- and gauge-invariant) operators of a given
dimension $d$ will be denoted $\acal_d.$  The number of linearly
independent operators is called the dimension $\dim\{\acal_d\}$ of
$\acal_d.$ To reduce the number of operators further requires the
equivalence theorem.

\section{Classification by Means of the Equivalence Theorem}\label{sec:equiv}

In this section, we will state the equivalence theorem and review
without proof some of the properties of equivalence relations,
equivalence classes, and quotient spaces.  We won't pause to provide
an illustration here, but we apply these concepts to a simple Yukawa
model in \ref{sec:yukawa}. In the next section, we shall marry these
ideas with the loop order to select the most phenomenologically useful
basis.

Among the operators in $\acal_d$, certain linear combinations
take the form
\beq\label{eq:kcal}
\sum_\phi U_\phi \frac{\delta S_0}{\delta \phi}+h.c.,
\eeq
where $U_\phi$ is some polynomial in the fields and their covariant
derivatives. (We implicitly include fermions and vector bosons in the
sum.) Here, $S_0$ is the action associated with the Lagrangian
$\lcal_0,$ \eqn{eq:efflag}, so ${\delta S_0}/{\delta \phi}=0$ represent
the classical EoM. The polynomials $U_\phi$ can depend on parameters
(masses and couplings) from $\lcal_0.$ $U_\phi$ must transform under
gauge transformations in such a way that the right-hand side is
gauge-invariant. Operators satisfying \eqn{eq:kcal} form a subspace
$\kcal_d\subset\acal_d$. (It is obviously sufficient to find operators
of the form $\sum_\phi U_\phi {\delta S_0}/{\delta \phi},$ since one can
always add its Hermitian conjugate.)

The equivalence theorem may be stated as follows:  Operators in
$\kcal_d,$ \ie, linear combinations of the form of \eqn{eq:kcal}, make
no contribution to S-matrix elements\footnote{See \cite{Politzer:1980me}. 
See also, \eg, \cite{Arzt:1993gz} and references therein.}. Thus, operators of this
type may be omitted in the construction of the effective Lagrangian,
\eqn{eq:efflag}. This is a very powerful result  that, as we shall see,
substantially simplifies $\lcal_{\rm eff}$ and reduces the number of
operators $\ocal_i$ required.  

Two $d$-dimensional operators $\ocal$ and $\ocal'$ are defined to be
{\it equivalent} if $(\ocal \!-\! \ocal')\!\in\!\kcal_d$.  In that case,
$\ocal$ and $\ocal'$ give the same contributions to observables.  For
this reason, the condition that their difference satisfies \eqn{eq:kcal}
is sometimes referred to as the ``on-shell" constraint. It can easily be
seen that this satisfies the requirements of an equivalence
relation~\cite{alg}. Therefore, every operator $\ocal$ may be associated
with a distinct set $[\ocal]$ of operators with which it is
equivalent.\footnote{There is no requirement that the operators under
consideration be monomials.}. The set $[\ocal]$ is called the {\it
equivalence class} associated with the operator $\ocal$.  The   
equivalence relation uniquely partitions the original set $\acal$ 
into distinct subsets.  The quotient space~\cite{alg} 
$\mcal_d\equiv\acal_d/\kcal_d$ consists of
the collection of all such equivalence classes.   

The equivalence relation allows one to replace the usual notion of
linear independence by inequivalence. Two operators whose difference
vanishes ``on-shell" can be regarded as identical for the purpose of
constructing $\lcal_{\rm eff}$, \eqn{eq:efflag}.  The way to state this
formally is to regard the quotient space $\mcal_d$ as a complex vector
space, with the null vector being identified with $[\kcal_d].$ The
number of inequivalent classes is the dimension of $\mcal_d,$ There is
a classic result~\cite{alg} that $\dim\{\mcal_d\} = \dim\{\acal_d\} -
\dim\{\kcal_d\}.$ Choosing one operator from each equivalence class
corresponds to selecting a basis set in which to express the higher
dimensional operators $\ocal_i$ in $\lcal_{\rm eff}$.
 
We may form the
union $\acal$ of all such operators $\acal\equiv \cup_d\, \acal_d,$ as
well as $\kcal\equiv \cup_d\, \kcal_d$  and $\mcal\equiv \cup_d
\mcal_d$. Of course, the dimensions of $\acal,\kcal,$ and $\mcal$ are
(denumerably) infinite. Since the elements of the quotient space $\mcal$
are the unique equivalence classes themselves, there is no ambiguity in
selecting a basis for $\mcal.$  Unfortunately, we do not know how to
calculate S-matrix elements starting directly from the equivalence
classes.  It is somewhat analogous to calculations with gauge fields. 
We know the results are gauge-invariant, but we must choose a gauge in
order to perform calculations.

A second, equivalent, way to approach this is to say that an operator
$\ocal$ from a set $\{\ocal_i\}$ is {\it redundant} if it is equal to a
finite linear combination of other operators, up to terms that vanish
on-shell, \ie, 
\beq\label{eq:equiv2} 
\ocal -\sum_i \kappa_i \ocal_i =
\sum_\phi U_\phi \frac{\delta S_0}{\delta \phi}. 
\eeq 
Starting from $\acal_d,$ there will be a minimum number of operators,
none of which are redundant. These are the operators that must be
included in the effective Lagrangian, \eqn{eq:efflag}.

To illustrate, consider simple $\vp^4$ theory 
\beq
\lcal_0=\half(\partial\vp)^2-V(\vp),\qquad V(\vp) = 
\half m^2\vp^2+\lambda\frac{\vp^4}{4}.
\eeq
The EoM is ${\delta \scal_0}/{\delta \vp} = -\partial^2\vp-V'(\vp).$
Considered as a model of low-energy physics, $V(\vp)$ has a discrete
$Z_2$ symmetry $\vp\to-\vp.$  We wish to consider possible extensions of
the theory as in \eqn{eq:model}, where $\lcal'$ has the same form as
$\lcal_0.$  Both are expressed in terms of renormalized fields, so that,
implicitly, $\Delta\lcal'$ contains appropriate counterterms.  In
general, we must assume that $\Delta\lcal'$ respects this $Z_2$ symmetry
since otherwise, the effective Lagrangian, \eqn{eq:efflag}, would
contain a term of the form $A\phi^3,$ where $A$ has dimensions of mass.
This would be required for renormalizability of \eqn{eq:efflag}.  Thus,
the higher dimensional operators in \eqn{eq:efflag} would involve only
even powers of $\phi,$ so the lowest order terms would have dimension
six.

This is already a non-trivial result, since it means that the first
corrections must be of $O(1/\Lambda^2)$ rather than $O(1/\Lambda).$ 
Considering dimension-six operators, we have $\{\vp^6,$
$\vp^2(\partial\vp)^2,$ $(\partial^2\vp)^2\}.$  Of these three, only one
is inequivalent, which we will take to be $\vp^6.$  To see this,
note that 
\beq
(\partial^2\vp)^2-V'(\vp)^2=-\left(\partial^2\vp-V'(\vp)\right)\frac{\delta S_0}{\delta \vp}
\eeq
Now $V'(\vp)^2 = (m^2\vp+\lambda\vp^3)^2 = m^4\vp^2 + 2m^2\lambda\vp^4 +
\lambda^2\vp^6.$  This shows that the operator $(\partial^2\vp)^2$ is
not inequivalent to the operators $\{\vp^6,\vp^4,\vp^2\}.$  

The occurrence of lower-dimensional operators $\{\vp^2, \vp^4\}$ in this
equivalence forces us into a slightly technical digression which however
has consequences for construction of all effective field theories. 
Obviously the occurrence of operators of lower dimension in such
relations is associated with light masses, here $m^2,$ or with
super-renormalizable couplings.  This complication may be dealt with in
several different ways, as elaborated in~\ref{sec:independent}.  The
upshot is that, for the purpose of determining inequivalence of
operators of a given dimension, one may simply ignore lower-dimensional
operators arising from application of the EoM.

Returning from this digression to classifying dimension-six operators,
we have shown that $(\partial^2\vp)^2$ is equivalent to 
$\vp^6.$  The same is true for $\vp^2(\partial\vp)^2.$ To see this,
first note that, after IBP, this operator is identical to $-\vp
\partial_\mu(\vp^2\partial^\mu\vp) \!=\!
-\vp^3\partial^2\vp\!-2\vp^2(\partial\vp)^2$, so that
$\vp^2(\partial\vp)^2\! \stackrel{IBP}{=}\!\! -\vp^3\partial^2\vp/3.$
Then we may use the EoM to replace $\partial^2\vp$,
\beq
 \vp^3\partial^2\vp +\vp^3V'(\vp)=-\vp^3\frac{\delta S_0}{\delta \vp}.
\eeq
Ignoring masses, the second term is simply $\lambda\vp^6.$  Therefore,
the only inequivalent correction of dimension six to the
effective Lagrangian can be taken to be $c_\vp\vp^6,$ as was claimed.

This simple example illustrates that one can continue this process of
eliminating redundant operators until one reaches a certain minimum
number of a given dimension.  We call this set a {\em basis;} note that
the linear (in $Q$)  nature of the relation \eqn{eq:equiv2} implies
that the {\em number} of elements in a basis is invariant. It also
suggests that, generally, it is easiest to use the EoM to remove
operators having multiple (covariant) derivatives. This strategy has
other benefits that will be explained subsequently.

Although the choice of basis operators for $\mcal_d$ is arbitrary, for
present purposes, we will want to select them with reference to their
order in the loop expansion, as will be discussed in the next section.

\section{Trees, Loops, \& Choice of Basis}\label{sec:basis}

Returning to \eqn{eq:model}, let us consider all possible
extensions $\Delta\lcal'(\phi_\ell,\phi_h)$ of a theory described by
$\lcal'(\phi_\ell).$  We wish to consider models having particles heavy
with respect to the particles described by $\lcal'.$  If not strongly
interacting, these too are expected to take the form of effective field
theories, which implies that the most relevant terms in $\Delta\lcal'$
will have dimensions four or less.  We will argue that the generic form
of such theories can be delineated without further assumptions and,
correspondingly, the leading observable corrections to $\lcal'$
discussed without reference to a specific underlying theory.  Further,
there is a way of selecting basis sets $\{ \ocal_i \}$ for
\eqn{eq:efflag} that, while not unique, are optimal, in a sense
described below.

Models will have fields representing possible scalars, fermions, and
vectors\footnote{While gravity could be included, we assume it is
irrelevant for the applications we have in mind, such as to LHC data.}. 
These must be consistent with the gauge symmetries associated with
$\lcal',$ and possibly with discrete or global symmetries as well.  The
interactions of the heavy fields with the light ones in $\lcal'$ can
only take the form of scalar self-interactions, Yukawa-like interactions
of scalars and fermions, and interactions with vectors as dictated by
the gauge symmetries.  As a first approximation, the most important of
these will usually be tree diagrams, so all the corresponding vertices
must contain at least one light field and one heavy field.  Using this,
one may write down all possible vertices consistent with the symmetries
of $\lcal'$ and draw all possible tree diagrams having external light
particles and internal heavy particles.  As an illustration, this was
done for the SM in ref.~\cite{Arzt:1994gp}; ({\it cf.} Fig.~3 in this
reference.)

From such diagrams, one may work out which operators $\ocal_i$ of a
given dimension $d$ may arise in the effective Lagrangian,
\eqn{eq:efflag}, from tree diagrams\footnote{Depending on the context, one may instead consider
restricted extensions of the theory, such as those having only
additional fermions or those having only a single additional heavy
vector boson or those respecting baryon-number conservation.  The
discussion here may be applied to such restricted cases as well.}. These operators form a
subset  of
the complete set of operators of a given dimension $d,$ and the vector
space spanned by them form a subspace of the space $\acal_d$ of all
operators of dimension $d.$  This subspace will be referred to as the
Potential-Tree-Generated (PTG) operators of dimension $d$ and denoted as
 $\acal_d^{PTG}.$   In practice, one will often only be interested in
the corrections of lowest order, operators of dimension five or 
six, although there are exceptions\footnote{See, \eg, 
\cite{delAguila:2012nu} and references therein.}.

Whether an operator is generated by a tree graph or only via loops
depends on the details of the underlying theory. The set $\acal_d^{PTG}
$ consists of those operators that are generated at tree-level in at
least one extension of the low-energy theory. In the case of the SM as
the low-energy theory, which describes nature well, there is no
guarantee that the BSM physics will in fact generate any of these at
tree level, hence our use of the qualifier ``potential".

How should one go about selecting a basis set for $\acal$?  In order to
deal with a finite number of operators, we shall focus on PTG operators
of a fixed dimension $d.$ Defining $[\acal_d^{PTG}]$ to be the set of
equivalence classes associated with all these operators, one may
consider its intersection with the quotient space
$\mcal_d\cap[\acal_d^{PTG}].$  For any equivalence class for which the
intersection is non-empty, it behooves us to select a basis operator for
that class from among operators in the intersection, because the PTG
operators generically have larger coefficients than the LG operators
\footnote{We are not the first to suggest this criterion.  
See~\cite{basis2}.}. 
Stated more physically, even though an equivalence class may contain
both operators from trees and from loops in the underlying theory,
without knowing the correct BSM dynamics, we ought to allow that the
coefficients are as large as potentially possible, since there are types
of BSM physics that do produce these operators at tree 
level.  Put yet
another way, if a LG operator were selected as the representative of
this equivalence class, experimental evidence for its presence may be
misinterpreted as being associated with a much lower threshold of new
physics than if it were a PTG operator. We call the equivalence classes
spanned by these classes $\mcal_d^{PTG}.$ If there remain any
equivalence classes for which the intersection is empty, then it is fair
to call these classes loop-generated (LG) $\mcal_d^{LG}$, since they
cannot arise from a tree graph in {\em any} extension of $\lcal'$.
Phenomenologically, this implies that the coefficients $c_i^{PTG}$ of
the dim\-ension-d basis operators of $\mcal_d^{PTG}$ take the form
$f_i/\Lambda^{d-4},$ with dimensionless constants $f_i\sim O(1)$ in all
models where these operators are in fact tree-generated\footnote{As
mentioned earlier, it may be that a naturalness requirement or other
constraint could influence expectations or interpretations.  {\it E.g.},
if, because of an enhanced chiral symmetry, each SM fermion mass 
vanishes in the limit of vanishing Yukawa couplings, then
$f_i$ for operators contributing to corrections to Yukawa
couplings must be proportional to a power of the SM fermion masses.}. On the
other hand, for operators from $\mcal_d^{LG},$ the coefficients may be
assumed to be  $f_i\sim O(1/16\pi^2)$ for {\em any} underlying theory.

For a simple example of how these concepts can be applied in practice,
we again refer to \ref{sec:yukawa}, where we classify operators as PTG
or LG and discuss the choice of basis operators for a simple Yukawa
model.  We want to move on to discuss the SM, but there are a couple of
caveats to keep in mind:  (1)~The equivalence of two operators is a
property of the low-energy theory alone.  It is possible for the
underlying theory to generate $ \ocal $ and not $ \ocal'$, even if the
two operators are equivalent.  (2)~It is possible for LG and PTG
operators to be equivalent.
This last point might appear odd, given the difference in the operator 
coefficients. To understand this better it is worth emphasizing that
what the equivalence theorem says is that the observable effects of one operator
can be mimicked by those of another one, it does not provide any insights
on the structure of the underlying physics. If 
$\ocal_{\rm PTG},~\ocal_{\rm LG} $ are equivalent PTG and LG operators
and are in fact generated by some heavy dynamics such that the first appears at
tree-level, the equivalence theorem does not imply that the effects of 
$ \ocal_{\rm LG} $ are promoted to tree-level status. Instead it says that
explicit calculation will show that the particles responsible for
generating $ \ocal_{\rm PTG} $ will provide deviations from the SM at tree-level,
while those that responsible for $ \ocal_{\rm LG} $ will generate small
additive corrections {\em as long as the typical energies are below $ \Lambda $}.
Above this scale, the PTG diagrams will exhibit resonances (in the 
appropriate channel) while  LG graphs will contain
the effects associated with 
unitarity cuts -- of course, in this regime the effective Lagrangian approach will no longer be applicable.

Usually, one would be satisfied seeking effects from the lowest
dimensional operators that may occur, but with some effort, one could do
even better.  From these classes $\mcal_d^{LG},$ one could identify all
operators that may arise at one-loop order in some underlying theory. 
If these did not exhaust all these equivalence classes, it would make
sense to define subspaces $\mcal_d^{1LG},$ with others arising in higher
order.  Phenomenologically, this is only interesting to do if one could
measure some S-matrix elements that could not arise from an underlying
theory until two-loops or higher.  This situation does in fact occur,
for example, in the calculation of electric dipole moments~\cite{cpv} in
certain models of CP-violation.

The different choices of basis operators will give different Green's
functions in general, even though each choice yields the same S-matrix
elements. One must be careful not to omit any basis operators without
good reason, since their contributions to Green's functions can be very
different.  An incomplete basis set may lead to spurious relations among
observables.  

\section{SM Operators of Dimension Six}

To put the preceding prescription into practice, we have classified all
the operators of dimension six for the SM, assuming the absence of
right-handed neutrinos and baryon- and lepton-conservation.  (If these
are violated, we assume it is at a much higher scale, as suggested by
limits on proton decay and on searches for lepton-violation~\cite{pdg}.)
The application of the equivalence theorem to the SM was originally
performed in a now-classic paper by Buchm\"uller \&
Wyler~\cite{Buchmuller:1985jz}  (BW.) It has been shown recently by
Grzadkowski \etal~\cite{Grzadkowski:2010es} (GIMR) that the BW ``basis"
in fact still contained redundant operators.  We will use the GIMR
basis, which, for easy reference, is reproduced in part in
\ref{sec:tables}, omitting the baryon-violating operators.  

The equivalence relations are summarized in \ref{sec:classes}.  If one
wishes to understand the implications for the equivalence classes, one
may simply replace each operator $\ocal$ by its corresponding
equivalence class  $[\ocal]$, and replace equivalence, $\sim$, by
equality, $=$, as was illustrated in \ref{sec:yukawa}.

To emphasize a point made in the previous section, we note that
\eqn{weom2} is an example of an equivalence relation involving both PTG
and LG operators.  One can show that $(D^\mu W_\mn^I)^2$ is a LG
operator, however every operator on the \rhs of that equivalence
relation is a PTG operator. This also illustrates that a particular sum
of PTG operators can be equivalent to a LG operator\footnote{Although
this has been asserted earlier in~\cite{Ellison:1998uy}, the example
given there is not correct.}. 

For our purposes, the specific formulae are less important than their
implications for the choice of basis operators. We have analyzed the
GIMR basis operators and determined which are PTG and which are LG.
Their basis can be classified as follows\footnote{The type denoted
$\psi^4$ includes all four-fermion operators. It turns out all are PTG,
regardless of their Lorentz or chiral structure.}:
\vskip-10mm
\bea
\label{ptgtype}{\rm PTG~types:} && \vp^6, \vp^4D^2, \psi^2\vp^3, \psi^2\vp^2D, \psi^4. \\
\label{lgtype}{\rm LG~types:} && X^3, X^2\vp^2, \psi^2 X\vp.
\eea
\vskip-2mm
By definition, no basis operator can be written as a linear combination
of other basis operators.  Even though 
certain linear combinations of PTG operators can be equivalent to a LG
operator, the basis chosen in~\cite{Grzadkowski:2010es} is optimal 
in the following sense: {\em none of their LG basis operators 
requires replacement by a PTG
operator,} \ie, the equivalence class of each LG basis operator contains
no PTG operators.  To establish this, one must examine each
equivalence relation in \ref{sec:classes} in which a given LG operator
occurs to determine whether there are any PTG operators in that class.  
The proof is by exhaustion; one need only examine each equivalence
relation in \ref{sec:classes} in which a given LG operator occurs and
verify that there are no PTG basis operators in that class.  It seems
the strategy of eliminating as many (covariant) derivatives as possible
has this unanticipated benefit.

\section{Triple Vector Boson Corrections}

As an application of the preceding, we return to the topic of
corrections to triple-gauge-boson (TGB) couplings from BSM physics.  In
earlier work~\cite{Arzt:1994gp}, we showed that corrections to triple
vector boson couplings do not arise in tree approximation in {\it any}
extension of the SM. In so doing, we compared operators arising from all
possible tree diagrams with the basis operators elaborated by
B\"uchmuller and Wyler (BW) \cite{Buchmuller:1985jz}. There are two
things we must do to update that analysis. First, as mentioned earlier,
it has been shown recently~\cite{Grzadkowski:2010es} that the basis of
\cite{Buchmuller:1985jz} contains redundant operators, so we should
determine whether this alters any of the conclusions of 
\cite{Arzt:1994gp} or \cite{Ellison:1998uy}.  Second, according to the
discussion above, it is not sufficient to complete the analysis using
one particular choice of basis operators, since a LG operator may be
equivalent to a linear combination of PTG operators.  Actually, this
second issue is the easier, since it was addressed in the preceding
section. Once we narrow down to the basis choice in
\cite{Ellison:1998uy}, then following the argument in
section~\ref{sec:basis}, it is not hard to see that none of their LG
operators is equivalent to a linear combination of PTG operators.

Returning to the first issue, the basis operators
in~\cite{Buchmuller:1985jz} that are redundant have been reviewed in
Section~3 of~\cite{Grzadkowski:2010es}, to which we refer for further
details.  Of the BW operators that affect the TGB couplings,
${\cal O}_\phi^{(1)}\!\equiv \! (\phi^\dag\phi)(D_\mu\phi)^\dag D^\mu\phi $
is
redundant, but it is also a PTG operator that is equivalent to a linear
combination of other PTG operators, so that is not a problem. Similarly,
the elimination of some four-fermion operators by Fierz transformations
is irrelevant to the TGB corrections.  None of the redundant operators
identified in ~\cite{Grzadkowski:2010es} having covariant derivatives on
fermion fields can contribute to the TGB couplings.   Thus, dropping the
redundant operators from the BW ``basis" has no effect on the analysis
of the TGB couplings in \cite{Arzt:1994gp}.

Therefore, none of the more recent developments vitiates the conclusions
in~\cite{Arzt:1994gp}.  In any extension of the SM, corrections to the
TGB couplings are suppressed by at least one-loop order.  Thus, assuming
that the BSM dynamics is decoupling, the magnitude of the coefficients
should only  be a few parts per thousand of the SM gauge coupling.

Note that if a process is not PTG, then whether the
corresponding dimension-six operator should be retained or
ignored depends not only on the on the process but also whether
the process is tree- or loop-generated in the SM. If it is SM
tree-generated (SMTG,) a BSM LG is probably going to be too
small to observe, but if the process is SM loop-generated
(SMLG), then a BSM loop operator (BSMLG) is only suppressed by a
factor of $\epsilon=v^2/\Lambda^2,$ just like a process that is
SMTG having PTG operators from BSM.

It is ironic in this context that the dominant production mechanism for
the production of the Higgs boson at LHC is gluon fusion (which is a
SMLG amplitude).  Similarly, one of the most easily identified decay
channels is $H\to\gamma\gamma,$  which is SMLG as well.  We shall
discuss this further elsewhere \cite{ew}.

Finally, note that the same is true of quartic vector boson
couplings, {\it viz.}\/ the only dimension-six operators
contributing to these are LG. Indeed, since they are associated
with the same operators as modifications of the
triple-vector-boson couplings, the magnitude of the two are
correlated.

Of course, for both triple and quartic couplings, it is possible
that other HDO, such as dimension-eight PTG operators, could be
more important than dimension-six LG operators, depending on the
scale of BSM physics. The exploration of these implications can
await experimental evidence for any deviations from the SM.

\section{Conclusion}\label{sec:conclude}

The purpose of this paper was to discuss how one might use the
equivalence relation not simply to establish the inequivalent operators
but also to determine the equivalence classes of operators. This is
similar to the demonstration of independence
in~\cite{Grzadkowski:2010es}, but we needed to make the equivalence
classes explicit in order to ascertain which among them contain PTG
operators; which, only LG operators; and which, both types.  We advocated
choosing a basis with the maximum number of PTG operators so that we
most easily and reliably interpret fits to data.  Even if no new physics
is indicated by experimental results, this strategy allows for a figure
of merit to be assigned and to infer the likely scale $\Lambda$ above
which we are ignorant.  With further restrictions on the BSM model, such
as supposing certain other global symmetries obtain, one may draw other
inferences.

We applied this to a classification of the dimension-six basis operators
of the SM, concluding that none of the LG basis operators
in~\cite{Grzadkowski:2010es} requires replacement by a PTG operator.  We
also revisited the triple-gauge boson (TGB) couplings in the present
light, reaffirming our earlier results~\cite{Arzt:1994gp} that
corrections to the TGB couplings do not arise at tree level.

In a certain sense, this classification scheme is renormalization group invariant, as can be deduced from earlier work~\cite{Einhorn:2001kj}.  
Notice first that the beta functions for the coefficients of the HDO's can be determined from Feynman rules involving only the basis set.  Although redundant operators may arise in this or another equivalence class, one may use the EoM to rewrite any such operator in terms of a sum of basis operators.  Under renormalization, operator mixing can occur with, for example, a PTG vertex contributing to the running of the coupling constant (coefficient) of a LG operator.  However, this does not invalidate our conclusions.  Consider a dimension-six LG operator.  If we simply connect two external legs, we get a  renormalization of a dimension-four operator, which we have agreed to absorb into the renormalized couplings of the SM.  So we must insert at least one SM vertex to obtain another dimension-six operator.  Thus, the evolution of the coupling constant multiplying a LG operator can involve a PTG vertex, giving a correction equal to the PTG vertex times a product of SM couplings, times at least one  loop-factor of $1/16\pi^2$.  This is at most of the same order as the original LG coupling and so does not upset the relative magnitude of the coefficients of PTG and LG operators.

It would be helpful if analytical methods could be found for identifying
PTG and LG operators for any effective field theory.  At present, the
only method we know is to delineate all vertices and to analyze all
potential tree graphs to determine which give rise to PTG operators.

There are obviously many opportunities to analyze data using these new
insights. In a related paper \cite{ew}, we apply this scheme to the production 
and decay of the SM Higgs boson.

\section{Acknowledgments}

We would like to thank the authors of ref.~\cite{Grzadkowski:2010es} for permission to use their tables.  MBE wishes to thank J.~R.\ Espinosa for discussions.  The research of one of us (MBE) was supported in part by the National Science Foundation under Grant No. NSF PHY11-25915.

\appendix

\section{\!\! Equivalence and Operators of Lower Dimension}\label{sec:independent}
The purpose of this appendix is to argue that lower dimensional
operators that occur in equivalence relations may be ignored because
they simply provide renormalizations of couplings already extant in
lower orders of the  effective field theory.  Such operators arise with
coefficients proportional to powers of superrenormalizable couplings in
$\lcal_0$ or to masses of light fermions and bosons.   This complication
may be dealt with in several different ways.  One may simply absorb such
terms into the coefficients of the lower dimensional operators in
$\lcal_{eff},$ or, perhaps even more simply, one may assume that the
lower dimensional operators are expressed in terms of  renormalized
fields and couplings, so that in addition to the terms
$\textstyle{\sum}c_i\ocal_i$, the interaction contains counterterms that
cancel  any corrections to operators already appearing in $\lcal_{eff}.$
Referring to the example at the end of  section~\ref{sec:efflag}, we
can say, \eg, that $\{(\partial^2\vp)^2, \vp^6\}$ are not
inequivalent dimension-six operators, with the understanding
that the equivalence may tacitly include operators of lower dimension as
well\footnote{One must keep in mind that this prescription does not take
into account quantities such as $m^2/\Lambda^2,$ whose size must be
compared to those of various renormalized couplings to which they may
contribute.  This is similar to the issue of {\it naturalness,}\/ which
transcends renormalizability.}.

The easiest way to deal with this complication in general is to
temporarily set to zero all masses or superrenormalizable couplings in
$\lcal_0.$  Then complete the classification of higher dimensional
operators with the understanding that, when these parameters are
restored, there will arise operators of lower dimension to be canceled by
appropriate counterterms.  This is the prescription that we will always
follow, so that it is as if operators of a given dimension do not mix
with operators of lower dimension under this equivalence relation.  (The
only caveat is that one must be sure that all such lower-dimensional
operators actually have been included.)

This technique is in harmony with mass-independent methods of
renormalization, such as dimensional regularization and minimal
subtraction, the standard method for treating gauge theories.  In this
way, renormalization of effective field theories remains consistent with
multiplicative renormalization.  Operators only mix with other operators
of the same dimension.  

\section{\!\! Simple Yukawa Model}\label{sec:yukawa}

In this appendix, we shall apply the formalism to determine the equivalence classes for a simple model.
Consider a Lagrangian with a scalar field $ \phi $ and a fermion $ \psi $:
\beq\label{eq:yukawa}
\lcal = \half ( \partial\phi)^2 - \half m^2 \phi^2  - \frac\lambda4 \phi^4+ \bar\psi ( i \!\! \not\!\partial - u \phi) \psi
\eeq
this has a Z$_2$ chiral symmetry $ \phi \to - \phi, \psi \to \gv \psi $.
 This symmetry rules out a Dirac mass $m_\psi\bar\psi \psi$. This
Lagrangian is also parity  invariant, $\psi\to \gamma^0\psi,$
$\phi\to+\phi.$  

The equation of motion are $ \ecal_\phi =0 ,~ \ecal_\psi = 0 $ where
\bea
\ecal_\phi &\equiv& \square\,\phi + \lambda \phi^3 + u \bar\psi\psi  \\
\ecal_\psi &\equiv& ( i \!\! \not\!\partial - u \phi) \psi
\eea

Considering extensions of this theory, what properties shall we require?
 We will assume that it respects the Z$_2$ chiral symmetry; otherwise,
we would expect the low energy theory to have a Dirac mass.  Whether it
must conserve parity is less clear, especially given experience with the
differences between electromagnetism and the electroweak theory.  For
simplicity, we will assume parity is conserved as well\footnote{If we
were describing a physical system rather than a simple model, these
would of course be testable assumptions.}.

Because of the Z$_2$ chiral symmetry, there are no dimension-five
operators.  It turns out that there are then seven types of
dimension-six operators\footnote{Ignore their designations as (PTG) or
(LG) for now.}:
\beq
\begin{array}{cllc}
1. & \phi^6: & \ocal_1 \equiv \phi^6 & (PTG) \cr
2. & \phi^4 \partial^2: & \ocal_2 \equiv \phi^3 \square\, \phi & (PTG)\cr
3. & \phi^3 \psi^2: & \ocal_3 \equiv \phi^3 \bar\psi\psi & (PTG) \cr
4. & \phi^2 \partial^4: & \ocal_4 \equiv ( \square\,\phi)^2  & (LG) \cr
5. & \phi^2 \psi^2 \partial: & \ocal_5 \equiv \phi^2 \bar\psi i \!\!\not\!\partial \psi,\  {\rm and}\  \ocal_5^\dagger &  (PTG)\cr
6. & \phi \psi^2 \partial^2:& \ocal_6\! \equiv\! (\square\,\phi) ( \bar\psi \psi) ;\, \ocal_7 \!\equiv\! ( \overline{i\!\!\not\!\partial \psi} \, i\!\!\not\!\partial \psi) \phi  ; & (LG) \cr &&\,\ocal_8 \!\equiv\! \phi ( \bar\psi \square \, \psi),\, {\rm and}\  \ocal_8^\dagger & (LG) \cr
\end{array}
\eeq
Finally, we have operators of the form $\psi^4$:
$ (\bar \psi \Gamma_a \psi)(\bar \psi
\Gamma^a \psi) $ with $ \Gamma_a \equiv \mati,$ $\gv,$ $\gamma_\mu,$
$\gv\gamma_\mu,$ $ \sigma_{\mu\nu} $, that we refer to as $S,~ P,~ V,
~A,~T$. 
The Fierz relations imply
\bea
2 T \!\!\!\!&=&\!\!\!\! -6 S + T - 6 P \!\!\!\then\!\!\! T =
-6 (S + P) \\ 2V \!\!\!\!&=&\!\!\!\! - 2S + V + A+2P
\!\!\!\then\!\!\!   A = V + 2 S - 2 P,\hskip5mm
\eea
\vskip-3mm
\ni so there are only three independent four-fermion operators, which we may choose as
\vskip-3mm
\beq
7.\ \ \ocal_9 \equiv S \ (PTG), \qquad  \ocal_9' \equiv P \  (PTG), \qquad  
\ocal_{10} \equiv V \  (PTG).
\eeq

The equivalence relation is defined as
\vskip-3mm
\beq 
\ocal \sim \ocal' \quad \Leftrightarrow \quad \ocal -  \ocal' =0 ~{\rm when}~ \ecal_\phi=\ecal_\psi =0.
\eeq
Then
\beq
\begin{array}{ll}
\ocal_2 \sim - \lambda \ocal_1 - u \ocal_3 & \ocal_4\sim -\lambda \ocal_2-u \ocal_6\cr
\ocal_5 \sim u \ocal_3 & \ocal_6 \sim - \lambda \ocal_3 - u \ocal_9 \cr
\ocal_7 \sim u^2 \ocal_3 & \ocal_8 \sim - u^2 \ocal_3
\end{array}
\eeq
From these, it follows that $\ocal_4 \sim \lambda^2 \ocal_1 + 2 \lambda
u \ocal_3 + u^2 \ocal_9,$ so that is not an independent relation.  There
are also the operators $\ocal_9^\prime$ and $\ocal_{10}$ which do not
appear in these relations, and so are inequivalent.

For the quotient space, we can simply replace in the preceding equations
 $\ocal_i\to[\ocal_i]$ and $\sim\,\to\,=$: 
\beq
\begin{array}{ll}
[\ocal_2] = - \lambda [\ocal_1] - u [\ocal_3] & [\ocal_4]= -\lambda [\ocal_2]-u [\ocal_6]\cr
[\ocal_5] = u [\ocal_3] & [\ocal_6] = - \lambda [\ocal_3] - u [\ocal_9] \cr
[\ocal_7] = u^2 [\ocal_3] & [\ocal_8] = - u^2 [\ocal_3]
\end{array}
\eeq
These six constraints among nine equivalence classes suggest that we
have at most 3 distinct equivalence classes among them.   With the benefit of 
hindsight, we may choose
them to be, \eg, $[\ocal_1],$ $[\ocal_3],$ and $[\ocal_9].$ The other 6
therefore are related; indeed,  
\beq\label{classes}
\begin{array}{ll}
[\ocal_2] = - \lambda [\ocal_1] - u [\ocal_3] & [\ocal_4]=\lambda^2 [\ocal_1] + 2 \lambda u [\ocal_3] + u^2 [\ocal_9] \cr
[\ocal_5] = u [\ocal_3] & [\ocal_6] = - \lambda [\ocal_3] - u [\ocal_9] \cr
[\ocal_7] = u^2 [\ocal_3] & [\ocal_8] = - u^2 [\ocal_3]
\end{array}
\eeq
In addition, there are the two single element equivalence classes
$[\ocal_9^\prime]$ and $[\ocal_{10}],$ which are inequivalent to others.
Thus, there are 5 independent equivalence classes, so there must be 5
independent basis operators.

Which five basis operators shall we choose?  Obviously, the  equivalence
relations have established that $\ocal_9'$ and $\ocal_{10}$ must be
chosen as basis operators, since their associated equivalence classes
contain only a single element.  For the remaining three equivalence classes,
according to our general approach, we want to consider the relations
among equivalence classes and, whenever possible, choose PTG operators.

Suppose we have an extension of this model described as in
\eqn{eq:model}
whose heavy particles generate (some of) the effective
operators $Q_i$.
We will assume that the kinetic energy terms for $\phi_\ell$ and
$\phi_h,$ which implicitly include fermions and gauge fields,
and that the
quadratic mass matrices contain no mixing terms\footnote{If
that is not the case, one can redefine the fields by performing
a global orthogonal (if real) or unitary (if complex)
transformation of the fields to bring it to this form. Of
course, one may have to adopt a renormalization procedure to
sustain the absence of such terms.} $\propto\phi_\ell \phi_h.$
In the absence of a quadratic mixing, the vertex structure of possible
extensions is like that of the SM, which we have already
discussed~\cite{Arzt:1994gp}. An examination of Fig.~3 in this reference 
leads to the classification of the eleven dimension-six operators\footnote{The
Hermitian conjugates, $\ocal_5^\dag$ and $\ocal_8^\dag,$ are not counted
separately.} as (PTG) or (LG), as we have denoted in the equations above. 
There are seven PTG operators and four LG
operators. The relations \eqn{classes} show that the equivalence classes
associated with all four LG operators, $[\ocal_4],$ $[\ocal_6],$
$[\ocal_7],$ $[\ocal_8],$ may be expressed in terms of classes
associated with PTG operators.  So we want to choose a basis from among
the PTG operators, {\it viz.,} $\ocal_1,$ $\ocal_3,$ and $\ocal_9.$ The
remaining two basis operators, $\ocal_9'$ and $\ocal_{10}$, are also PTG
operators.  Therefore, we can choose a complete set of 5 basis operators
that are all PTG.

If this model were actually testable experimentally and precise
measurements were carried out that determined or placed limits on the
five independent coupling constants $c_i$ associated with these
operators, then one may be able to infer some hints about the underlying
extended theory.
For example, if there were evidence for the presence
of the operator $\ocal_{10},$ but, with the same precision, no evidence
for $\ocal_9$ or $\ocal_9'$, one would be strongly motivated to search
for a heavy vector boson.  Assuming its gauge couplings, parameterized
by $f_{10},$ were $O(1),$ we could infer an approximate upper limit on
its mass.

\pagebreak

\section{\!\! Dimension-Six Basis Operators for the SM\protect\footnote{These tables are taken from~\cite{Grzadkowski:2010es}, by permission of the authors.}.} \label{sec:tables}

\vskip-5mm
\begin{table}[h]
\renewcommand{\arraystretch}{1.5}
\begin{tabular}{||c|c||c|c||c|c||}
\hline \hline
\multicolumn{2}{||c||}{$X^3$\ \bf{(LG)} } & 
\multicolumn{2}{|c||}{$\vp^6$~ and~ $\vp^4 D^2$ \bf{(PTG)} } &
\multicolumn{2}{|c||}{$\psi^2\vp^3$\ \bf{(PTG)} }\\
\hline
$Q_G$                & $f^{ABC} G_\mu^{A\nu} G_\nu^{B\rho} G_\rho^{C\mu} $ &  
$Q_\vp$       & $(\vp^\dag\vp)^3$ &
$Q_{e\vp}$           & $(\vp^\dag \vp)(\bar l_p e_r \vp)$\\
$Q_{\wt G}$          & $f^{ABC} \wt G_\mu^{A\nu} G_\nu^{B\rho} G_\rho^{C\mu} $ &   
$Q_{\vp\Box}$ & $(\vp^\dag \vp)\raisebox{-.5mm}{$\Box$}(\vp^\dag \vp)$ &
$Q_{u\vp}$           & $(\vp^\dag \vp)(\bar q_p u_r \tvp)$\\
$Q_W$                & $\eps^{IJK} W_\mu^{I\nu} W_\nu^{J\rho} W_\rho^{K\mu}$ &    
$Q_{\vp D}$   & $\left(\vp^\dag D^\mu\vp\right)^\star \left(\vp^\dag D_\mu\vp\right)$ &
$Q_{d\vp}$           & $(\vp^\dag \vp)(\bar q_p d_r \vp)$\\
$Q_{\wt W}$          & $\eps^{IJK} \wt W_\mu^{I\nu} W_\nu^{J\rho} W_\rho^{K\mu}$ &&&&\\    
\hline \hline
\multicolumn{2}{||c||}{$X^2\vp^2$\ \bf{(LG)}} &
\multicolumn{2}{|c||}{$\psi^2 X\vp$\ \bf{(LG)}} &
\multicolumn{2}{|c||}{$\psi^2\vp^2 D$\ \bf{(PTG)} }\\ 
\hline
$Q_{\vp G}$     & $\vp^\dag \vp\, G^A_{\mu\nu} G^{A\mu\nu}$ & 
$Q_{eW}$               & $(\bar l_p \sigma^{\mu\nu} e_r) \tau^I \vp W_{\mu\nu}^I$ &
$Q_{\vp l}^{(1)}$      & $(\vpj)(\bar l_p \gamma^\mu l_r)$\\
$Q_{\vp\wt G}$         & $\vp^\dag \vp\, \wt G^A_{\mu\nu} G^{A\mu\nu}$ &  
$Q_{eB}$        & $(\bar l_p \sigma^{\mu\nu} e_r) \vp B_{\mu\nu}$ &
$Q_{\vp l}^{(3)}$      & $(\vpjt)(\bar l_p \tau^I \gamma^\mu l_r)$\\
$Q_{\vp W}$     & $\vp^\dag \vp\, W^I_{\mu\nu} W^{I\mu\nu}$ & 
$Q_{uG}$        & $(\bar q_p \sigma^{\mu\nu} T^A u_r) \tvp\, G_{\mu\nu}^A$ &
$Q_{\vp e}$            & $(\vpj)(\bar e_p \gamma^\mu e_r)$\\
$Q_{\vp\wt W}$         & $\vp^\dag \vp\, \wt W^I_{\mu\nu} W^{I\mu\nu}$ &
$Q_{uW}$               & $(\bar q_p \sigma^{\mu\nu} u_r) \tau^I \tvp\, W_{\mu\nu}^I$ &
$Q_{\vp q}^{(1)}$      & $(\vpj)(\bar q_p \gamma^\mu q_r)$\\
$Q_{\vp B}$     & $ \vp^\dag \vp\, B_{\mu\nu} B^{\mu\nu}$ &
$Q_{uB}$        & $(\bar q_p \sigma^{\mu\nu} u_r) \tvp\, B_{\mu\nu}$&
$Q_{\vp q}^{(3)}$      & $(\vpjt)(\bar q_p \tau^I \gamma^\mu q_r)$\\
$Q_{\vp\wt B}$         & $\vp^\dag \vp\, \wt B_{\mu\nu} B^{\mu\nu}$ &
$Q_{dG}$        & $(\bar q_p \sigma^{\mu\nu} T^A d_r) \vp\, G_{\mu\nu}^A$ & 
$Q_{\vp u}$            & $(\vpj)(\bar u_p \gamma^\mu u_r)$\\
$Q_{\vp WB}$     & $ \vp^\dag \tau^I \vp\, W^I_{\mu\nu} B^{\mu\nu}$ &
$Q_{dW}$               & $(\bar q_p \sigma^{\mu\nu} d_r) \tau^I \vp\, W_{\mu\nu}^I$ &
$Q_{\vp d}$            & $(\vpj)(\bar d_p \gamma^\mu d_r)$\\
$Q_{\vp\wt WB}$ & $\vp^\dag \tau^I \vp\, \wt W^I_{\mu\nu} B^{\mu\nu}$ &
$Q_{dB}$        & $(\bar q_p \sigma^{\mu\nu} d_r) \vp\, B_{\mu\nu}$ &
$Q_{\vp u d}$   & $i(\tvp^\dag D_\mu \vp)(\bar u_p \gamma^\mu d_r)$\\
\hline \hline
\end{tabular}
\caption{\sf Dimension-six operators other than the four-fermion ones.\label{tab:no4ferm}}
\end{table}
\pagebreak
\begin{table}[h]
\begin{center}
{\bf All are PTG.}
\end{center}
\renewcommand{\arraystretch}{1.5}
\begin{tabular}{||c|c||c|c||c|c||}
\hline\hline
\multicolumn{2}{||c||}{$(\bar LL)(\bar LL)$} & 
\multicolumn{2}{|c||}{$(\bar RR)(\bar RR)$} &
\multicolumn{2}{|c||}{$(\bar LL)(\bar RR)$}\\
\hline
$Q_{ll}$        & $(\bar l_p \gamma_\mu l_r)(\bar l_s \gamma^\mu l_t)$ &
$Q_{ee}$               & $(\bar e_p \gamma_\mu e_r)(\bar e_s \gamma^\mu e_t)$ &
$Q_{le}$               & $(\bar l_p \gamma_\mu l_r)(\bar e_s \gamma^\mu e_t)$ \\
$Q_{qq}^{(1)}$  & $(\bar q_p \gamma_\mu q_r)(\bar q_s \gamma^\mu q_t)$ &
$Q_{uu}$        & $(\bar u_p \gamma_\mu u_r)(\bar u_s \gamma^\mu u_t)$ &
$Q_{lu}$               & $(\bar l_p \gamma_\mu l_r)(\bar u_s \gamma^\mu u_t)$ \\
$Q_{qq}^{(3)}$  & $(\bar q_p \gamma_\mu \tau^I q_r)(\bar q_s \gamma^\mu \tau^I q_t)$ &
$Q_{dd}$        & $(\bar d_p \gamma_\mu d_r)(\bar d_s \gamma^\mu d_t)$ &
$Q_{ld}$               & $(\bar l_p \gamma_\mu l_r)(\bar d_s \gamma^\mu d_t)$ \\
$Q_{lq}^{(1)}$                & $(\bar l_p \gamma_\mu l_r)(\bar q_s \gamma^\mu q_t)$ &
$Q_{eu}$                      & $(\bar e_p \gamma_\mu e_r)(\bar u_s \gamma^\mu u_t)$ &
$Q_{qe}$               & $(\bar q_p \gamma_\mu q_r)(\bar e_s \gamma^\mu e_t)$ \\
$Q_{lq}^{(3)}$                & $(\bar l_p \gamma_\mu \tau^I l_r)(\bar q_s \gamma^\mu \tau^I q_t)$ &
$Q_{ed}$                      & $(\bar e_p \gamma_\mu e_r)(\bar d_s\gamma^\mu d_t)$ &
$Q_{qu}^{(1)}$         & $(\bar q_p \gamma_\mu q_r)(\bar u_s \gamma^\mu u_t)$ \\ 
&& 
$Q_{ud}^{(1)}$                & $(\bar u_p \gamma_\mu u_r)(\bar d_s \gamma^\mu d_t)$ &
$Q_{qu}^{(8)}$         & $(\bar q_p \gamma_\mu T^A q_r)(\bar u_s \gamma^\mu T^A u_t)$ \\ 
&& 
$Q_{ud}^{(8)}$                & $(\bar u_p \gamma_\mu T^A u_r)(\bar d_s \gamma^\mu T^A d_t)$ &
$Q_{qd}^{(1)}$ & $(\bar q_p \gamma_\mu q_r)(\bar d_s \gamma^\mu d_t)$ \\
&&&&
$Q_{qd}^{(8)}$ & $(\bar q_p \gamma_\mu T^A q_r)(\bar d_s \gamma^\mu T^A d_t)$\\
\hline\hline 
\multicolumn{2}{||c||}{$(\bar LR)(\bar RL)$ and $(\bar L R)(\bar L R)$}\\
\cline{1-2}
$Q_{ledq}$ & $(\bar l_p^j e_r)(\bar d_s q_t^j)$ \\
$Q_{quqd}^{(1)}$ & $(\bar q_p^j u_r) \eps_{jk} (\bar q_s^k d_t)$ \\
$Q_{quqd}^{(8)}$ & $(\bar q_p^j T^A u_r) \eps_{jk} (\bar q_s^k T^A d_t)$ \\
$Q_{lequ}^{(1)}$ & $(\bar l_p^j e_r) \eps_{jk} (\bar q_s^k u_t)$ \\
$Q_{lequ}^{(3)}$ & $(\bar l_p^j \sigma_{\mu\nu} e_r) \eps_{jk} (\bar q_s^k \sigma^{\mu\nu} u_t)$ \\
\cline{1-2}
\cline{1-2}
\end{tabular}
\caption{\sf Four-fermion operators conserving baryon number. \label{tab:4ferm}}
\end{table}

\section{\!\! SM Equivalence Relations for Dimension-Six Operators}\label{sec:classes}

In the following we use the following notation:
$\vp$ denotes the SM scalar isodoublet; $u,\, d,\, e$ denote
the right-handed up, down and charged lepton fields, while
$ l,\, q $ denote the lepton and quark left-handed doublets.
 $G_\mu^A,\
, W_\mu^I$ and $B_\mu$ correspond to the 
$\su3_c $, $\su2_L $ and $\ui_Y$ gauge fields and
$g_s,\, g,\, g'$  the corresponding gauge couplings.
$ \Gamma_{u,d,e}$ denote the Yukawa-coupling matrices,
$\lambda$ the Higgs self-coupling constant;
$p,r,s,t$ are family indices and
 $i,j,k$ denote the $\su2_L$ indices for $l$ and $q$.
 $T^A = \lambda^A/2$ denote the $\su3_c$ generators in the
fundamental representation, $ f^{ABC} $ the corresponding
structure constants; and $ \varepsilon{ij},\,
\varepsilon^{IJK} $ the completely antisymmetric tensors
in two and three indices.
All the operators  on the left-hand-side below, except the first, are LG; for ease of reading
we group them according to the field content.
%
\paragraph{Scalars and vectors:}
\beq 
\label{phi4D2} (\vp^\dag\vp)(\vp^\dag D^2\vp)\sim\lambda Q_{\vp}\!-\! \Gamma_{\!e}^\dag Q_{e\vp}^\dag\!-\!\Gamma_{\!d}^\dag Q_{d\vp}^\dag\!-\!\Gamma_{\!u} Q_{u\vp},
\eeq

\beq \label{vpjt2b}
 (\vpjt)^2\!\sim\! 3 Q_{\vp\Box}\!-\!4\lambda Q_\vp\!+\!2\left(\Gamma_{\!e}\,Q_{e\vp}\!+\!
 \Gamma_{\!d}\,Q_{d\vp}\!+\!\Gamma_{\!u}\,Q_{u\vp}\!+\!h.c.\right).
\eeq
\bea \label{phi2d4}
(D^2\vp)^\dag  D^2\vp  &\sim & \lambda^2 Q_\vp - 
2\lambda\big(\Gamma_{\!e}Q_{e\vp}\!+\Gamma_{\!d}Q_{d\vp}\!+\Gamma_{\!u}Q_{u\vp}\!+\!h.c.\big)+
\Gamma_{\!e}^\dag\Gamma_{\!e}  Q_{\ell e}
+\Gamma_{\!d}^\dag\Gamma_{\!d}Q_{\ell d} + \cr &&\!+
\Gamma_{\!u}^\dag\Gamma_{\!u}Q_{\ell u}
\!+\!\big(\Gamma_{\!d}^\dag\Gamma_{\!e}Q_{\ell edq}\!+ 
\Gamma_{\!e}\Gamma_{\!u}Q_{\ell equ}^{(1)}\!+
\Gamma_{\!d} \Gamma_{\!u} Q_{quqd}^{(1)}\!+\!h.c. \big).
\eea
\beq \label{equivB}
(\vpj)\p_\nu B^\mn \!\sim\! \frac{g'}{2}\big(4Q_{\vp D}\!+\!Q_{\vp\Box}\big) 
\!-
g'\!\sum_s\!\big({\textstyle\frac{1}{2}}Q^{(1)}_{\vp\ell}\!+\!\ocal_{\vp e}\!-\! {\textstyle\frac{1}{6}}Q^{(1)}_{\vp q}\!+\!{\textstyle\frac{1}{3}}Q_{\vp d}\!-\!{\textstyle\frac{2}{3}}Q_{\vp u}\big)_{\!s}. 
\eeq
\bea \label{equivW}
(\vpjt)D_\nu W^{I\mn} &\sim&  g \Big[\textstyle{\f32} Q_{\vp\Box}\!-\!2\lambda Q_\vp  
+ \sum_s \big(Q_{\vp\ell}^{(3)}+Q_{\vp q}^{(3)}\big)_{\!s} + \cr &&
+ \big(\Gamma_{\!e}\,Q_{e\vp}\!+\Gamma_{\!d}\,Q_{d\vp}\!+\Gamma_{\!u}\,Q_{u\vp}+h.c.\big) 
\!\Big]
\eea
%
\paragraph{Only vectors:}
\bea \label{beom2}
(\p_\mu B^\mn)^2  &\sim& g^{\prime\,2}\Big[Q_{\vp D}\!+\! \textstyle\f14 Q_{\vp\Box}\!+\!
 \sum_s \big(\!-\!\frac{1}{2} Q^{(1)}_{\vp\ell}\!-\!Q_{\vp e}\!+\! \frac{1}{6}Q^{(1)}_{\vp q}\!-\!\frac{1}{3} Q_{\vp d}+ 
\cr &&
\!+\! \textstyle\frac{2}{3} Q_{\vp u}\big)_{\!s}\!+\!\sum_{ps}\Big(\f14 Q_{\ell\ell}^{(1)}\!+\!\f1{36} Q_{qq}^{(1)}\!-\!\f16 Q_{\ell q}^{(1)} \!+\! Q_{ee}\!+\! \f19 Q_{dd}^{(1)}\!+\!\f49 Q_{uu}^{(1)}\!+\!  \f23 Q_{ed}-\! 
\cr&&
\textstyle \!-\!\f43 Q_{eu}\!-\!\f49 Q_{ud}^{(1)} \!+\! Q_{\ell e}\!+\!   \textstyle \f13 Q_{\ell d}\!-\!\f23 Q_{\ell u}\!-\! \f13 Q_{qe} \!-\!\f19 Q_{qd}^{(1)}\!+\!\f29 Q_{qu}^{(1)}\Big)_{\!ps}\, \Big].
\eea
\bea \label{weom2}
(D^\mu W^I_\mn)^2 & \sim& \f{g^2}{4}\Big[\!-\!4\lambda Q_\vp+3 Q_{\vp\Box} +2\big(\Gamma_{\!e}\,Q_{e\vp}\!+\Gamma_{\!d}\,Q_{d\vp}\!+\Gamma_{\!u}\,Q_{u\vp}+h.c.\big)+
\cr&&
+2\,\textstyle{\sum_s}\big(Q_{\vp\ell}^{(3)}+Q_{\vp\ell}^{(3)}\big)_{\!s}
+\textstyle{\sum_{ps}}\big(Q_{\ell\ell}^{(1)}+Q_{qq}^{(3)}+2 Q_{\ell q}^{(3)}\big)_{\!ps}\, \Big].
\eea
\bea \label{geom2}
\big(D^\mu G^A_\mn\big)^2 &\sim& g_s^2\, \textstyle{\sum_{st}}\Big[\f14\big(Q^{(3)}_{qq}\!+\!Q^{(1)}_{qq}\big)_{\!stts}\!+
\textstyle \f12\big(Q_{dd}^{(8)}\!+\!Q_{uu}^{(8)} \big)_{\!stst} -\! \cr&& \qquad
\!-\textstyle\f16\big(Q^{(1)}_{qq}\!+\!Q^{(1)}_{qd}\!+\!Q^{(1)}_{qu} \big)_{\!st}\!+2\big(Q_{qd}^{(8)}\!+\!Q_{qu}^{(8)}\!+\!Q_{ud}^{(8)}\big)_{\!st}\Big].
\eea
%
%
\paragraph{Fermions and vectors:}
\bea \label{psi2D3}
 \label{ellell2}
(\overline{i\D\ell})i\,\Dfbs (i\,\D \ell)\!\!\!\!&\sim&\!\!\!\!
\Gamma_e^\dag\Gamma_e\Big[\Gamma_e Q_{e\vp} \!+\! \Gamma_e^\dag Q_{e\vp}^\dag\!+\! Q_{\vp e}\Big], \cr &&\cr
\label{ee2}
 (\overline{i\D e})i\,\Dfbs (i\D e)\!\!\!\!&\sim&\!\!\!\!
\Gamma_e\Gamma_e^\dag\Big[\Gamma_e Q_{e\vp}+
\Gamma_e^\dag Q_{e\vp}^\dag \!-\!\textstyle{\half}\big(Q_{\vp\ell}^{(3)}+Q_{\vp\ell}^{(1)}\big)\Big],\hskip2mm  \cr &&\cr
\label{qq2}
 (\overline{i\D q})i\,\Dfbs (i\D q)\!\!\!\!&\sim&\!\!\!\!
 \Big[ \Gamma_u\Gamma_u^\dag\Gamma_u\, Q_{u\vp}\!+\!
  \Gamma_d \Gamma_d^\dag \Gamma_d\, Q_{d\vp}\!+\! h.c. \Big]\!+\! \cr 
  &~& 
 + 2\Big[\Gamma_u^\dag\Gamma_u\, Q_{\vp u}\!+\! \Gamma_d \Gamma_d^\dag\, Q_{\vp d} 
 + 2\Re \big(\Gamma_u^\dag\Gamma_d\, Q_{\vp u d}\big)\Big],\\ &&\cr
 \label{dd2}
(\overline{i\D d})i\, \Dfbs (i\D d)\!\!\!\!&\sim&\!\!\!\!
 \Gamma_d^\dag\Gamma_d\Big[\Gamma_d Q_{d\vp}\!+\Gamma_d^\dag Q_{d\vp}^\dag\!-\!
 \textstyle{\half}\big(Q_{\vp q}^{(3)}+Q_{\vp q}^{(1)}\big)\Big],\cr &&\cr
 \label{uu2}
(\overline{i\D u})i\,\Dfbs  (i\D u)\!\!\!\!&\sim&\!\!\!\! \Gamma_u\Gamma_u^\dag\Big[\Gamma_u Q_{u\vp} \!+\Gamma_u^\dag Q_{u\vp}^\dag\!-\!\textstyle{\half}\big(Q_{\vp q}^{(3)}+Q_{\vp q}^{(1)}\big)\Big].\nnb 
\eea

\bea\label{octetequiv2}
 (\bar q \gamma^\mu T^A q)_p(D^\mu G^A_\mn) &\sim &  g_s \,
\textstyle{\sum_{s}}\big(Q_{qq}^{(8)}\!+Q_{qd}^{(8)}\!+Q_{qu}^{(8)}\big)_{\!ps},\cr
(\bar d \gamma^\mu T^A d)_p(D^\mu G^A_\mn) &\sim &
g_s\,\textstyle{\sum_{s}}\big(  Q_{qd}^{(8)}+ 
\textstyle{\half}Q_{dd}^{(1)} -\f16 Q_{dd}^{(1)}+Q_{ud}^{(8)}\big)_{\!ps}, \\
(\bar u \gamma^\mu T^A u)_p(D^\mu G^A_\mn) &\sim &
g_s\,\textstyle{\sum_{s}}\big(  Q_{qu}^{(8)}+Q_{ud}^{(8)}+
\textstyle{\half}Q_{uu}^{(1)} -\f16 Q_{uu}^{(1)}\big)_{\!ps}. \nnb
\eea

\bea
(\bar q \gamma^\mu \tau^I q)_p(D^\mu W^I_\mn) \!&\sim &\!  \f{g}{2}\big[
(Q_{\vp q}^{(3)})_p + \textstyle{\sum_{s}} \big(Q_{\ell q}^{(3)}+Q_{q q}^{(3)}\big)_{\!ps} \big] ,\cr
(\bar\ell \gamma^\mu \tau^I \ell)_p(D^\mu W^I_\mn) \!&\sim &\! 
\f{g}{2}\big[ (Q_{\vp \ell}^{(3)})_p  +  \textstyle{\sum_{s}}\big(
 Q_{\ell\ell}^{(3)}+ Q_{\ell q}^{(3)}\big)_{\!ps}\big].  
\eea
\bea \label{psi2DX}
(\bar\ell\gamma^\mu\ell)_p \p^\rho B_{\rho\mu} \!\!\!\!&\sim&\!\!\!\!
\frac{g'}{2}(Q_{\vp\ell})_p\!+\!
g'\textstyle{\sum_{s}}\big(\!\!-\!\frac{1}{2}Q_{\ell\ell}^{(1)}\!+\!\frac{1}{6}Q_{\ell q}^{(1)}\!-\!Q_{\ell e}\!-\!
\frac{1}{3}Q_{\ell d}\!+\!\frac{2}{3}Q_{\ell u}\big)_{\!ps}, \hskip5mm \nnb \\
(\bar q\gamma^\mu q)_p\p^\rho B_{\rho\mu} \!\!\!\!&\sim&\!\!\!\!
\frac{g'}{2}(Q_{\vp q})_p\!+\!
g'\textstyle{\sum_{s}}\big(\!\!-\frac{1}{2}Q_{\ell q}^{(1)}\!+\!\frac{1}{6}Q_{qq}^{(1)}\!-\!Q_{qe}\!-\!
\frac{1}{3}Q_{q d}^{(1)}\!+\!\frac{2}{3}Q_{q u}^{(1)}\big)_{\!ps}, \hskip5mm\nnb\\
(\bar e\gamma^\mu e)_p\p^\rho B_{\rho\mu} \!\!\!\!&\sim&\!\!\!\!
\frac{g'}{2}(Q_{\vp e})_p\!+\!
g'\textstyle{\sum_{s}}\big(\!\!
-\!\frac{1}{2}Q_{\ell e}\!+\!\frac{1}{6}Q_{qe}\!-\!Q_{ee}\!-\!
\frac{1}{3}Q_{de}\!+\!\frac{2}{3}Q_{ue}\big)_{\!sp}, \\
(\bar d\gamma^\mu d)_p\p^\rho B_{\rho\mu} \!\!\!\!&\sim&\!\!\!\!
\frac{g'}{2}(Q_{\vp d})_p\!+\!
g'\textstyle{\sum_{s}}\big(\!\!-\frac{1}{2}Q_{\ell d}\!+\!\frac{1}{6}Q_{qd}^{(1)}\!-\!Q_{ed}\!-\!
\frac{1}{3}Q_{d d}\!+\!\frac{2}{3}Q_{ud}^{(1)}\big)_{\!sp},\hskip5mm\nnb\\
(\bar u\gamma^\mu u)_p\p^\rho B_{\rho\mu} \!\!\!\!&\sim&\!\!\!\!
\frac{g'}{2}(Q_{\vp u})_p\!+\!
g'\textstyle{\sum_{s}}\big(\!\!-\frac{1}{2}Q_{\ell u}\!+\!\frac{1}{6}Q_{qu}^{(1)}\!-\!Q_{eu}\!-\!
\frac{1}{3}Q_{u u}\!+\!\frac{2}{3}Q_{du}^{(1)}\big)_{\!sp}, \hskip5mm\nnb
\eea  
\beq \label{psi2DX}
(\bar \ell \,\sigma^\mn  i\D \ell) B_\mn \!\sim\! 
\Gamma_{\! e} Q_{eB}, \hskip5mm
\qquad\qquad
(\bar e \,\sigma^\mn  i\D e) B_\mn \!\sim\!\
\Gamma_{\! e}^\dag Q_{eB}^\dag,
\eeq
\bea
(\bar q \,\sigma^\mn  i\D q) B_\mn  &\sim &
\,\Gamma_{\! d} Q_{dB}\!+\!\Gamma_{\! u} Q_{uB},\cr
(\bar d \,\sigma^\mn  i\D d) B_\mn &\sim & \Gamma_{\! d}^\dag Q_{dB}^\dag,\\
(\bar u \,\sigma^\mn  i\D u) B_\mn &\sim & \Gamma_{\! u}^\dag Q_{uB}^\dag.\nnb
\eea
\beq
(\bar \ell \,\sigma^\mn \tau^I i\D \ell) W_\mn^I \!\sim\! 
\Gamma_{\! e} Q_{eW}, 
\qquad\qquad
(\bar e \,\sigma^\mn \tau^I i\D e) W_\mn^I \!\sim\! 
\Gamma_{\! e}^\dag Q_{eW}^\dag, \hskip7mm
\eeq
\bea
(\bar q \,\sigma^\mn \tau^I i\D q) W_\mn^I &\sim&
\,\Gamma_{\! d} Q_{dW}\!+\!\Gamma_{\! u} Q_{uW}, \cr
(\bar d \,\sigma^\mn \tau^I i\D d) W_\mn^I &\sim&
\Gamma_{\! d}^\dag Q_{dW}^\dag,  \\
(\bar u \,\sigma^\mn \tau^I i\D u) W_\mn^I &\sim& 
\Gamma_{\! u}^\dag Q_{uW}^\dag. \nnb
\eea
\bea
(\bar q \,\sigma^\mn T^A i\D q) G_\mn^A &\sim&
\,\Gamma_{\! d} Q_{dG}\!+\!\Gamma_{\! u} Q_{uG}, \cr
(\bar d \,\sigma^\mn T^A i\D d) G_\mn^A &\sim&
\Gamma_{\! d}^\dag Q_{dG}^\dag, \\
(\bar u \,\sigma^\mn T^A i\D u) G_\mn^A &\sim&
\Gamma_{\! u}^\dag Q_{uG}^\dag.\nnb
\eea
%
\paragraph{Fermions, scalars and vectors:}
\bea \label{psi2D2phi}
(\bar \ell e)D^2\vp \!\!\!\!&\sim&\!\!\!\! 
\lambda Q_{e\vp}\!+\!\textstyle{\half}\Gamma_{\!e}^\dag Q_{\ell e}
\!-\!\Gamma_{\!d}^\dag Q_{\ell edq}\!+\!\Gamma_u Q_{\ell equ}^{(1)}\, , \cr
(\bar q d)D^2\vp \!\!\!\!&\sim&\!\!\!\! 
\lambda Q_{d\vp}\!-\!\Gamma_{\!e}^\dag Q_{\ell edq}^\dag\!+\!
\textstyle{\half}\Gamma_d^\dag Q_{qd}^{(1)}\!-\!\Gamma_u Q_{quqd}^{(1)}\, , \\
(\bar q  u)D^2\tvp \!\!\!\!&\sim&\!\!\!\! 
\lambda Q_{u\vp}\!-\!\Gamma_{\!e}Q_{\ell equ}^{(1)}{}^{\!\!\!\!\dag}-\!\Gamma_{\!d}Q_{quqd}^{(1)}
\!+\!\textstyle{\half}\Gamma_u^\dag Q_{qu}^{(1)}.\nnb
\eea
\bea \label{yukawa}
(\overline{i\D e}) \vp^\dag (i\D \ell)\!\!\!\!&\sim&\!\!\!
\Gamma_{\!e}^2 Q_{e\vp}, \cr
 (\overline{i\D d}) \vp^\dag (i\D q)\!\!\!\!&\sim&\!\!\!
\Gamma_{\!d}^2 Q_{d\vp}, \\
 (\overline{i\D u}) \tvp^\dag (i\D q)\!\!\!\!&\sim&\!\!\!
 \Gamma_{\!u}^2(\overline{q} \tvp)(\tvp^\dag \tvp) u\!=\!\Gamma_{\!u}^2 Q_{u\vp}. \nnb
\eea
\smallskip
\ni We now define $( {\cal O}_{\vp\psi})_{pr}\equiv (\vp^\dag iD_\mu\vp) (\bar \psi_p \gamma^\mu \psi_r),$
where $p,r$ are generation indices. 
\bea \label{psi2phi2D}
\hspace{-10mm}(i D_\mu\vp)^\dag(\bar e \gamma^\mu i\D \ell)\!\sim\!\Gamma_{\!e}\,  {\cal O}_{\vp e}^\dag ,\qquad\quad\ \, &~&\!\!\!\!\!\!\!\!
(i D_\mu\vp)(\bar \ell \gamma^\mu i\D e)\!\sim\!\textstyle{\half}\Gamma_{\!e}^\dag \big( {\cal O}_{\vp\ell}^{(1)}\!+\!{\cal O}_{\vp\ell}^{(3)}\big),\hskip10mm\nnb\\
\hskip10mm(i D_\mu\vp)^\dag(\bar d \gamma^\mu i\D q)\!\sim\!\Gamma_{\!d}\, {\cal O}_{\vp d}^\dag \!+\!
\Gamma_{\!u}\,Q_{\!\vp ud}^\dag, \!\!\!&~&\!\!\!\!\!\!\!\!
(i D_\mu\vp)(\bar q \gamma^\mu i\D d)\!\sim\!\textstyle{\half}\Gamma_{\!d}^\dag\big( {\cal O}_{\vp q}^{(1)}\!+\!{\cal O}_{\vp q}^{(3)} \big)\!,\hskip8mm\\
(i D_\mu\tvp)^\dag(\bar u \gamma^\mu i\D q)\!\sim\!\Gamma_{\!u}\,  {\cal O}_{\vp u}^\dag \!+\!\Gamma_{\!d}\,Q_{\vp ud},
\!\!\!&~&\!\!\!\!\!\!\!
(i D_\mu\tvp)(\bar q \gamma^\mu i\D u)\!\sim\!\textstyle{\half}\Gamma_{\!u}^\dag\big( {\cal O}_{\vp q}^{(1)}
\!+\!{\cal O}_{\vp q}^{(3)} \big).\hskip10mm\nnb
\eea 
\bea \label{phi2Dpsi2}
 (\vp^\dag\vp)(\bar \ell i\D \ell) &\sim& \Gamma_{\!e}Q_{e\vp}, \qquad\quad
(\vp^\dag\vp)(\bar e i\D e) \sim \Gamma_{\!e}^\dag Q_{e\vp}^\dag,\cr
(\vp^\dag\vp)(\bar q i\D q)&\sim & \Gamma_{\!d}Q_{d\vp}\!+\! \Gamma_{\!u}Q_{u\vp}, \\ 
(\vp^\dag\vp)(\bar d i\D d) &\sim &\Gamma_{\!d}^\dag Q_{d\vp}^\dag, \qquad\qquad
(\vp^\dag\vp)(\bar u i\D u) \sim   \Gamma_{\!u}^\dag Q_{u\vp}^\dag.\nnb
\eea
\beq \label{phi2Dpsi2vector}
(\vp^\dag \tau^I \vp)(\bar \ell \tau^I i\D \ell)\!\sim\! 
\Gamma_{\!e}\,Q_{e\vp},\qquad\qquad
(\vp^\dag \tau^I \vp)(\bar q \tau^I i\D q)\!\sim\! 
\Gamma_{\!d}\,Q_{d\vp}\!+\! \textstyle{\half}\Gamma_{\!u}Q_{u\vp}.
\eeq
\pagebreak

\end{document}